\documentclass[amsmath,amssymb,pra,aps,twocolumn,mathbbm,superscriptaddress]{revtex4}
\usepackage{amssymb}
\usepackage{graphicx}
\usepackage{times}
\usepackage{subfigure}
\usepackage{color}

\newcommand{\ket}[1]{\ensuremath{\left|{#1}\right\rangle}}
\newcommand{\bra}[1]{\ensuremath{\left\langle{#1}\right|}}

\begin{document}

\title{The quantum Rabi model in a superfluid Bose-Einstein condensate}

\author{S. Felicetti}
\address{Laboratoire Mat\' eriaux et Ph\' enom\`enes Quantiques, Sorbonne Paris Cit\' e, Universit\' e Paris Diderot, CNRS UMR 7162, 75013, Paris, France}
\author{G. Romero}
\address{Departamento de F\'isica, Universidad de Santiago de Chile (USACH),  Avenida Ecuador 3493, 9170124, Santiago, Chile}
\author{E. Solano}
\address{Department of Physical Chemistry, University of the Basque Country UPV/EHU, Apartado 644, E-48080 Bilbao, Spain}
\address{IKERBASQUE, Basque Foundation for Science, Maria Diaz de Haro 3, 48013 Bilbao, Spain}
\author{C. Sab\'{i}n}
\address{Instituto de F\'isica Fundamental, CSIC, Serrano 113-bis 28006 Madrid, Spain}


\begin{abstract}
We propose a quantum simulation of the quantum Rabi model in an atomic quantum dot, which is a single atom in a tight optical trap coupled to the quasiparticle modes of a superfluid Bose-Einstein condensate. This widely tunable setup allows to simulate the ultrastrong coupling regime of light-matter interaction in a system which enjoys an amenable characteristic timescale, paving the way for an experimental analysis of the transition between the Jaynes-Cummings and the quantum Rabi dynamics using cold-atom systems. Our scheme can be naturally extended to simulate multi-qubit quantum Rabi models. In particular, we discuss the appearance of effective two-qubit interactions due to phononic exchange, among other features.
\end{abstract}

\flushbottom
\maketitle
%
%
\thispagestyle{empty}

\section{Introduction} The engineering of light-matter interaction lies at the core of modern quantum science. Its development has allowed us to implement highly controllable quantum technologies where atomic or solid-state systems interact with confined modes of the electromagnetic field. Physical implementations include cavity quantum electrodynamics (QED) \cite{Walther2006,QuantumBook}, ultracold atoms \cite{IBloch2002}, trapped ions \cite{IonsReview}, hybrid systems \cite{HybridDevices}, and circuit QED \cite{Blais04,Wallraff04,Chiorescu04,DevoretSchoelkopf}. A turning point in the history of quantum technologies has been marked by the  achievement of the strong coupling (SC) regime~\cite{Haroche13,Wineland13}, where the interaction strength overcomes dissipation rates. In the SC regime, coherent quantum processes can be observed and even applied in quantum information tasks.   

When the light-matter interaction strength becomes comparable with the systems bare frequencies, the ultrastrong coupling (USC) regime~\cite{Ciuti05,Ciuti06} is reached,  where the system spectrum and dynamical features are fundamentally modified. The minimal model of light-matter interaction is called quantum Rabi model~\cite{Rabi1936} (QRM), and it consists of a two-level quantum system (qubit) coupled with a single bosonic mode. In spite of its apparent simplicity, the dynamics of the QRM in the USC regime cannot be solved analytically, and its eigenspectrum has been  obtained only recently~\cite{Braak2011}.
The USC regime has been experimentally observed using semiconductor quantum wells \cite{Anappara2009,Gunter2009,Todorov2010}, molecular cavity QED~\cite{Schwartz11,George16}
and superconducting quantum circuits~\cite{Bourassa2009,Niemczyk2010,Pol2010}. Recently, the latter technology allowed to couple matter excitations with a continuum of bosonic modes~\cite{Pol17}, or with extreme values of the coupling strength~\cite{Yoshihara17}, a condition called deep strong coupling (DSC) regime.
The growing interest on the USC regime is motivated not only by its fundamental features, but also by potential quantum-information applications, such as parity-protected quantum computation \cite{Nataf2011, FelicettiReports}, quantum information storage \cite{Kyaw2014a}, and ultrafast quantum information processing \cite{Romero2012}. 

Although exciting, natural implementations of systems in the USC regime are still very challenging and they are confined by fundamental limitations. However, using quantum-simulation schemes, the physics of the USC regime can be observed also in systems that do not naturally achieve this regime of interaction~\cite{Ballester12,Crespi12,Pedernales2015, Felicetti15, Langford16,Fedortchenko17,Felicetti17}. These techniques allow to explore all parameter regimes, to overcome fundamental limitations and to observe exotic interactions.
 In particular, different schemes have been proposed to implement the QRM using superconducting circuits~\cite{Ballester12}, quantum optical systems~\cite{Crespi12}, trapped ions~\cite{Pedernales2015} and cold atoms~\cite{Felicetti17}. An experimental simulation of the QRM has been recently implemented~\cite{Langford16}, using a digital quantum simulation scheme on superconducting quantum devices.

\begin{figure}[t] 
\centering
\includegraphics[width=\linewidth]{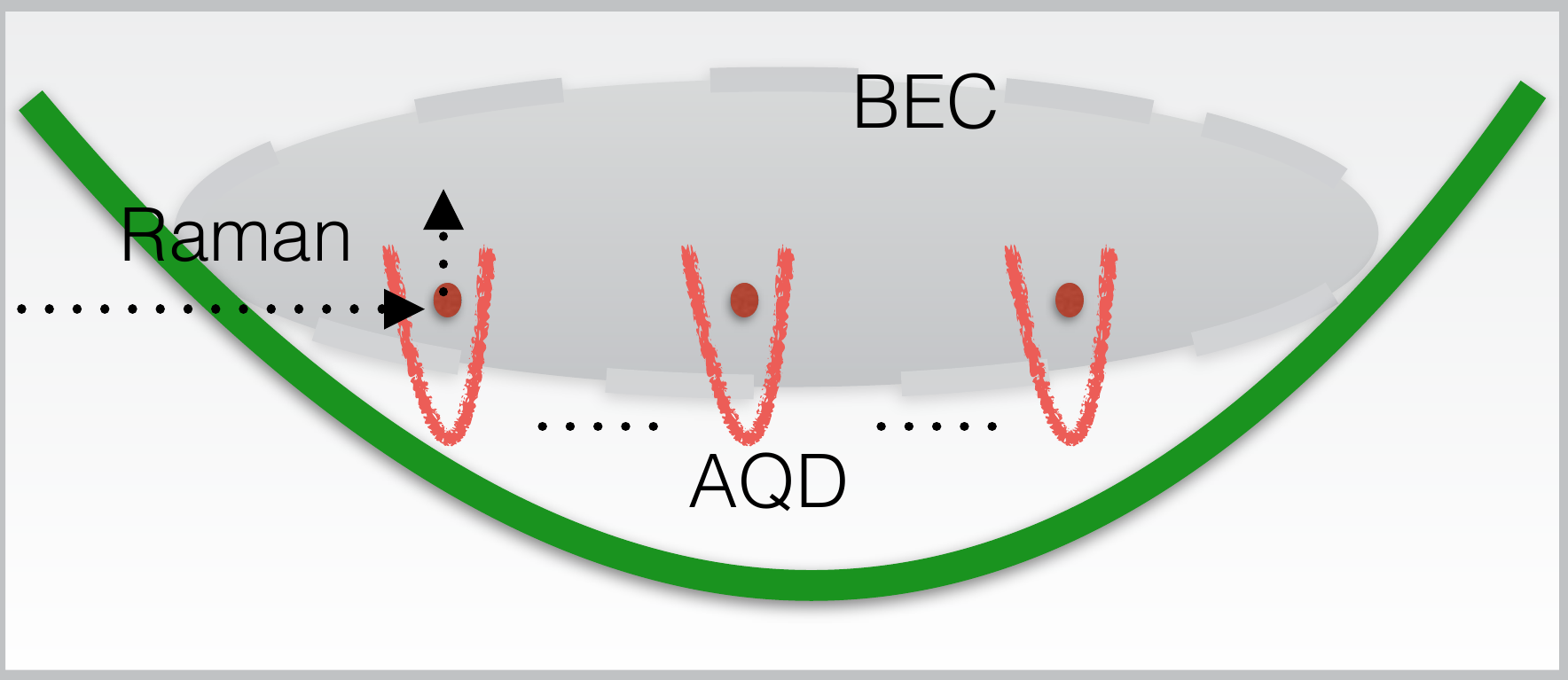}
 \caption{(Color online) Schematic of the setup. Several atomic quantum dots (AQD) are coupled to the BEC atomic cloud in the superfluid regime by a Raman laser, which generates a phonon-mediated interaction.}
\label{fig:setup}
\end{figure}

In this work, we propose an alternative setup for the quantum simulation of the quantum Rabi model in a cold atom system, see Fig. \ref{fig:setup}. The proposed scheme consists of a superfluid Bose-Einstein condensate (BEC) in a shallow confining trap, where an atomic quantum dot interacts with phononic quasiparticle modes of the BEC. All the parameters can be widely tuned, allowing to observe the transition between strong, ultrastrong and deep strong coupling regime. Furthermore, due to the slow speed of propagation of the quasiparticles, the characteristic frequencies of this setup can be as low as several $\operatorname{Hz}$, resulting in a favourable timescale for real-time control of the artificial matter-radiation interaction.  Therefore, our setup offers a promising testbed for the experimental analysis of the full quantum Rabi model (QRM)\cite{Rabi1936,Braak2011}. Moreover, our scheme allows to implement a controllable interaction between atomic quantum dots, mediated by dispersive coupling with phononic modes. In this case, the possibility to reach the USC regime provides an effective qubit-qubit interaction~\cite{Zueco2009} that is robust against thermal effects~\cite{Cardenas2016}.
\section{Quantum simulation scheme}

 The system here considered consists in a Bose-Einstein condensate (BEC) superfluid reservoir in a shallow confining trap and one or more two-level quantum systems (TLSs). One possible implementation of a TLS is an atomic quantum dot \cite{Recati2005, Orth2008,RecatiGinzburg}. The dot is created by applying a localized steep potential which traps atoms of a different hyperfine state and of the same atomic species as the BEC. We will use $a$ and $b$ to denote atoms in the BEC and the dot traps respectively. We consider the collisional blockade regime, requiring the interactions between atoms in the dot to be much larger than the interactions between atoms in the condensate reservoir $g_{bb}\gg g_{aa}$, so that the dot occupation number can only be 0 or 1, giving rise to a TLS.  Alternatively, in reference \cite{recatiII} the TLSs are created by loading cold atoms of a different species in an optical lattice consisting of double-well potentials where only single occupancy of each site is allowed.

We consider the BEC reservoir in the low temperature superfluid regime, in which the density of the condensate can be split into a background mean-field term and a quantum fluctuation operator $\Pi(\bf{x})$, that can be expanded in terms of Bogoliubov modes. For frequencies below a cutoff frequency $\omega_c$, given by the condensate healing length, these modes describe phononic excitations,

\begin{eqnarray}
  \Pi(x)=\sum_{k} \sqrt{N\omega_k}\left[e^{ikx}a_k +\mathrm{H.c.}\right].\label{field}
\label{field}
\end{eqnarray}
This field is described by a continuum of bosonic modes, with creation and annihilation Fock operators that satisfy the commutation relation $[a_k,a^{\dag}_{k'}]=\delta_{kk'},$ and a linear spectrum, $\omega_k = v|k|,$ where $v$ is the propagation velocity of the phononic excitations, given by $m v^{2}=\rho_{a} g_{aa}$ in a weakly interacting BEC. Here, $m$ and $\rho_{a}$ are the mass of bosons in the condensate and the condensate density, respectively. The coupling strength $g_{aa}$ is expressed as  $g_{aa}=4\pi\hbar^2\,a_{aa}/m$, where $a_{aa}$ the scattering length of atoms in the BEC. Therefore, the speed of sound in the BEC is proportional to the square root of the scattering length, which can be controlled by tuning an external magnetic field around a Feshbach resonance \cite{feshbach1,feshbach2, feshbach3}.  Finally, $N$ is a normalization constant $N=\hbar/(2\,V\,g_{aa})$, that depends on the condensate volume $V$. The dots are coupled to the condensate reservoir through a Raman transition of effective Rabi frequency $\Omega$ and detuning $\delta$ by means of external lasers, which give rise to a phonon-mediated interaction. 

Within the above approximations, the system is well described (see the derivation in \cite{Recati2005}) by a spin-boson like Hamiltonian, 
\begin{eqnarray}\label{spin-boson}
H &=& \left(\frac{\hbar\Omega_d}{2}\sigma^{z}+\left[-\delta'+\sum_{\textbf{k}}g\left(a_{\textbf{k}}+a_{\textbf{k}}^{\dagger} \right) \right]\frac{\hbar\sigma^{x}}{2}\right)+\nonumber\\ &+&\sum_{\textbf{k}}\hbar\omega_{\textbf{k}}a_{\textbf{k}}^{\dagger}a_{\textbf{k}} \, ,
\end{eqnarray}
 where $\sigma_{x}$, $\sigma_{z}$ are Pauli spin matrices.  The first term in Eq.(\ref{spin-boson}) is the free Hamiltonian of an effective qubit with energy gap $\hbar\,\Omega_d$. The frequency $\Omega_d$ is determined by the effective Rabi frequency $\Omega$ of the Raman transition and the number of atoms in the condensate.  The second term describes the coupling of the dot with the phononic field, whose Hamiltonian is given by the third term in Eq. (\ref{spin-boson}). The dot-phonon coupling is characterized by 
 \begin{equation}\label{eq:coupling}
g=\sqrt{\frac{N \omega_{k} }{\hbar^2 }}\left(g_{ab}-g_{aa}\right)\, .
\end{equation}
Here, $g_{ab}$ is the coupling strength between the dot and condensate atoms, given by $g_{ab}=4\pi\hbar^2\,a_{ab}/m$.  $\delta'$ is dependent on $g_{ab}$, $g_{aa}$, $\Omega$ and $\delta$ \cite{Recati2005}. As all the coupling strengths, $g_{ab}, g_{aa}$ and $g_{bb}$, can be tuned by Feshbach resonances, and we are able to consider the case $\delta'=0$. 
In this case, the interaction Hamiltonian for a single dot can be rewritten as
\begin{equation}
H_{I}= \hbar (g_{ab}-g_{aa})\Pi({\bf{x}}_a).\label{eq:interaction}
\end{equation}
With suitable boundary conditions for the condensate trap, e. g. hard-wall or box potentials \cite{condensatebox1,condensatebox2,condensatebox3}, the energy separation of the phononic modes can be large enough to ensure that each TLS is only effectively coupled to the mode with closest frequency to $\Omega_d$.  We denote this frequency by $\Omega_f$.
In a 3D condensate, taking $V=L^3$, $L= 10\operatorname{\mu m}$ and $v=10 \operatorname{mm /s}$ -- typical numbers in the literature-- , the frequency of the lowest-energy phonons is $\Omega_f\simeq 2\pi\times v/(2\,L)\simeq2\pi\times500\,\operatorname{Hz}$. 
For instance, we can consider a K-Rb mixture, where Rb is the atomic species of the condensate and K that of the impurities. Then, the scattering length $a_{\text{RbRb}}=a_{aa}=102\,a_0$ and $a_{KRb}=a_{ab}$ can be tuned near the Feshbach resonance from $a_{ab}\simeq a_{aa}$ to $a_{ab}>30\,a_{aa}$ \cite{KRb}. Then $g$ can be tuned within a broad range from $2 \pi\times5\operatorname{Hz}$ to $2 \pi\times\operatorname{100}\operatorname{Hz}$. In lower dimensions, the value of $g$ can be larger. For instance, if we consider a cylindrical quasi 1D BEC with tight radial confinement $L_{r}\ll L_{z}=L$,  we have to replace $V$ by $L$ inthe definition of $N$, and the couplings $g_{ij}$ by  $g^{(1D)}_{ij}=g_{ij}/L_r^2$ \cite{pitastring}. Therefore, we obtain $g^{(1D)}_{ij}=\lambda^{-1/2}\,g_{ij}$, where $\lambda= (L_{r}/L_{z})^2\ll 1$ is the aspect ratio of the condensate. With a typical value $\lambda\simeq 10^{-3}$ \cite{1Dbec}, we get $g_{ij}^{(1D)}\simeq 30\,g_{ij}$. 

The readout can be performed including additional atomic quantum dots coupled to the BEC in the dispersive regime. In this way, the effective qubit can be used to realize a quantum non-demolition measurement of the average number of phonons. Indeed, it has been shown that this setup can be used as a highly sensitive thermometer at ultralow temperatures, which in turn amounts to measure the average number of phonons in a field mode \cite{fedichev2003gibbons,thermometer}. Remarkably, a large number of independent atomic quantum dots can be used. Alternatively, an optical lattice loaded with cold atoms in the single occupancy regime would give rise to a similar model \cite{recatiII}.

It is worth mention that our setup is different from other possible implementation of the QRM in cold-atoms systems. In a spin-orbit coupled condensate \cite{socoup, socoup2} two-level systems are encoded in internal atomic hyperfine levels of the Raman-dressed condensate, while the bosonic mode refers to the atomic wavefunction in a harmonic trap, that is not to the quasiparticle modes. This configuration allows to observe impressive many-body phenomena, but it cannot be applied to implement few-body interactions in the USC regime.
A scheme to implement the QRM in cold atoms systems has also been proposed~\cite{Felicetti17}, where both the qubit and the bosonic mode are encoded in the motion of the atomic cloud in tailored optical lattices~\cite{diracweitz, diracweitz2}. This proposal was designed to selectively implement a novel region of parameters in the DSC regime that was previously unexplored, and which could not be accessed by other means. Finally, the statistics  of the quantum Rabi model energy levels could be reproduced by bosonic atoms trapped in a double-well potential together with a single-impurity atom~\cite{mumford2014impurity}.



\section{The quantum Rabi model} The above analysis shows that  the USC and DSC regimes of interaction can indeed be achieved. In the following, we will focus on the single-qubit case, and we will address the changes of the system dynamics as the interaction strength is increased.  In this case, the system dynamics is governed by the QRM according to the Hamiltonian
\begin{equation}
H_{\rm QRM} = \frac{\hbar\Omega_d}{2}\sigma^z + \hbar\Omega_fa^{\dag}a + \hbar g\sigma^x(a + a^{\dag}).
\label{RabiH}
\end{equation}
A critical issue in the proposed implementation is the cooling of the bosonic excitations. In order to observe Rabi oscillations between the TLS and the bosonic field, the latter must be initialized in the vacuum or in a few-phonon state. To this end, the initial temperature should be low enough to ensure that the average number of thermal excitations in the relevant range of frequencies is close to 0. At $500\, \operatorname{Hz}$, this number is below $10^{-1}$ at $T=10\, \operatorname{nK}$. Temperatures as low as $0.5\,\operatorname{nK}$ has been achieved in a BEC \cite{sciencewr}. We have performed numerical simulations in order to analyze interesting dynamics that could be implemented with feasible physical parameters, including the effect of finite-temperature initialization of the bosonic field. Simulations have been performed by direct diagonalization of the system Hamiltonian in the Fock basis. In order to obtain a finite-size system, a cut off $N<100$ has been imposed on the maximum number of photons allowed into the system. The dissipation in the bosonic mode with bare decay parameter $\gamma$ and temperature has been taken into account by means of the microscopic master equation \cite{Ciuti2009}. 

\subsection{Strong to Ultrastrong coupling transition}

First, let us consider the transition from the strong to the USC regime. When the coupling strength is small compared to the system frequencies, the quantum Rabi Hamiltonian (\ref{RabiH}) can be approximated via rotating-wave approximation (RWA) to the Jaynes-Cummings Hamiltonian
\begin{equation}
H_{\rm JC} = \frac{\hbar\Omega_d}{2}\sigma^z + \hbar\Omega_fa^{\dag}a + \hbar g( \sigma^+a +  \sigma^-a^{\dag}).
\label{JCH}
\end{equation}
In this case, the Hamiltonian respects a continuous $U(1)$ symmetry that preserves the total number of excitations $\hat{N}=a^{\dag}a+\sigma^+\sigma^-$. The system dynamics consists in coherent exchange of excitations between the TLS and the field, dubbed Rabi oscillations. As the coupling strength increases and the system reaches the USC regime, the full QRM of Eq.\eqref{RabiH} must be considered, and the continuous symmetry breaks down into a discrete $\mathbb{Z}_2$-symmetry, defined by the parity operator $\hat{P} = -\sigma^z e^{i\pi a^{\dag}a}$. Accordingly, the number of excitations is not conserved, and the parity symmetry sets selection rules for state transitions \cite{FelicettiReports,PolReports}. The transition from the SC to the USC regime entails fundamental modification to the system ground state, which is not the vacuum but contains virtual photons. As a result, the Jaynes-Cummings doublets of phonons and TLS excitations do not provide an intuitive description of the system dynamics.

\begin{figure}[] 
\centering
\includegraphics[width=\linewidth]{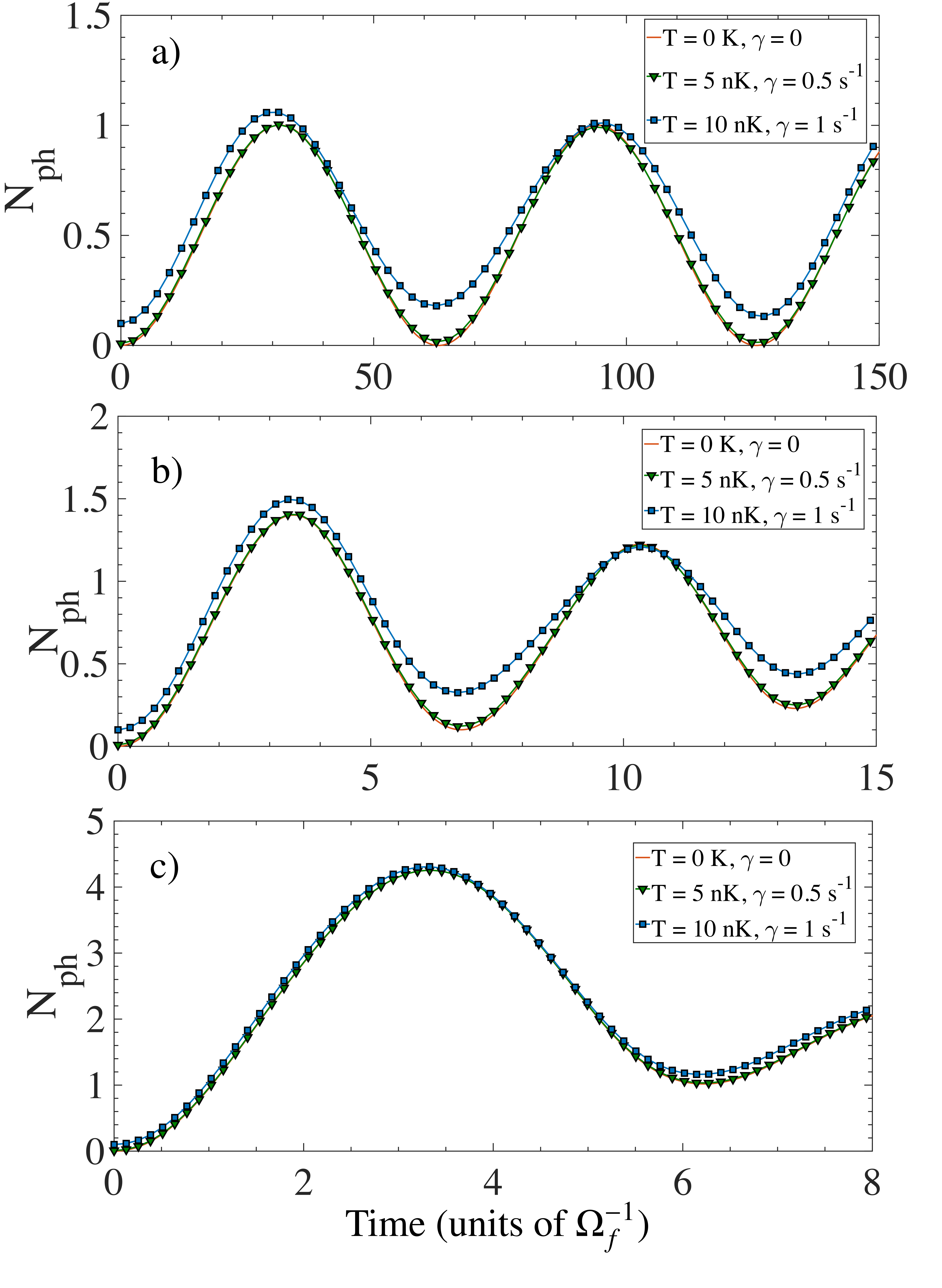}
\caption{(Color online) Average photon number along time evolution, for different values of the coupling strength. a) $g=0.05\ \Omega_f$, b) $g=0.5\ \Omega_f$, and c) $g= \Omega_f$. In all plots, the TLS is taken to be resonant with the bosonic mode $\Omega_d=\Omega_f$. For all cases, the continuous red line corresponds to $T= 0\ \text{nK}$ and $\gamma = 0\ \text{s}^{-1}$; for the inverted green triangles $T= 5\ \text{nK}$ and $\gamma = 0.5\ \text{s}^{-1}$; finally, for the blue squares $T= 10\ \text{nK}$ and $\gamma = 1\ \text{s}^{-1}$.}
     \label{Fig2}
\end{figure}

\begin{figure}[] 
\centering
\includegraphics[width=\linewidth]{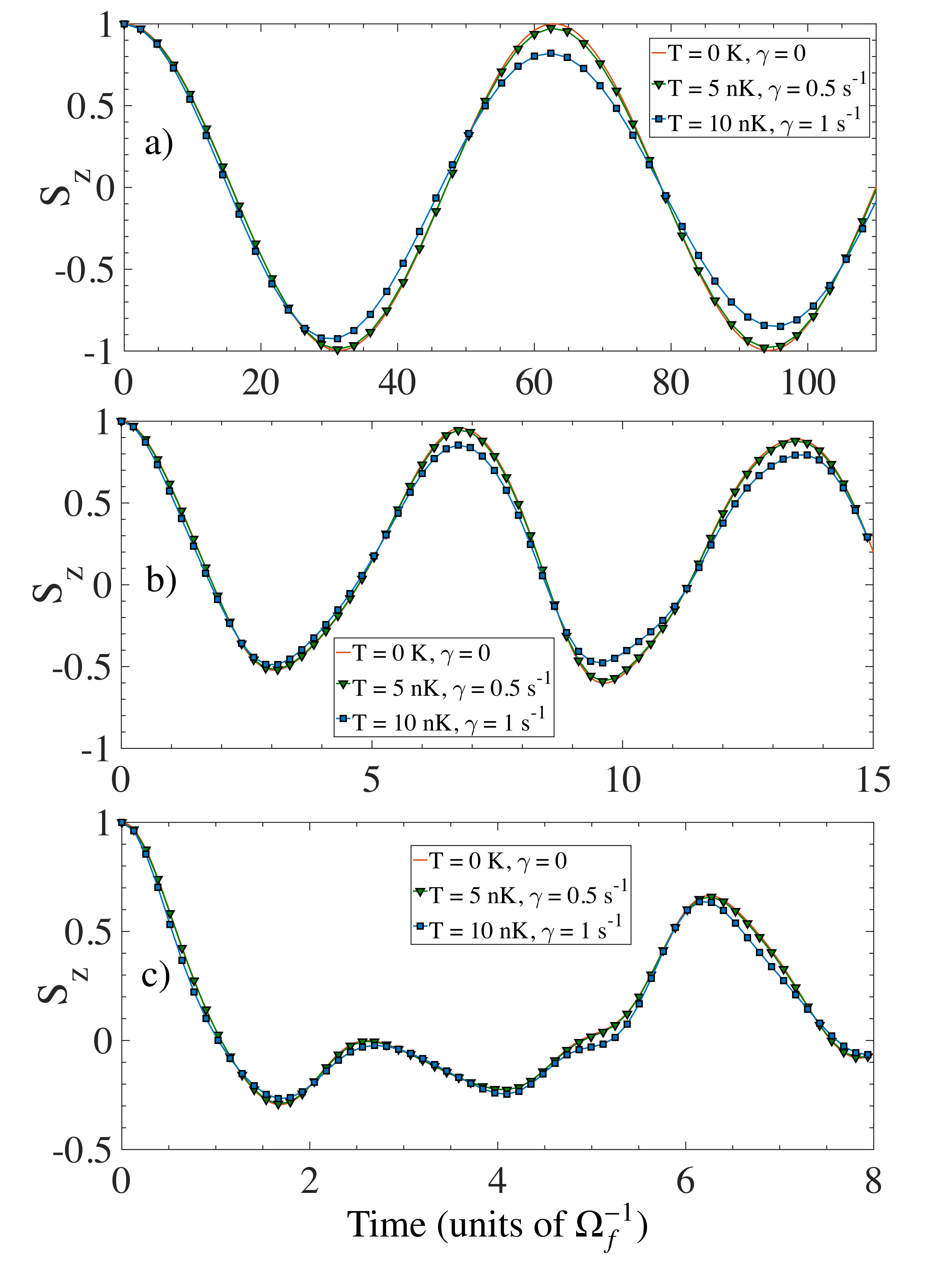}
    \caption{(Color online) Expected value of $\sigma_z$ along time evolution, for different values of the coupling strength. a) $g=0.05\ \Omega_f$, b) $g=0.5\ \Omega_f$, and c) $g= \Omega_f$. In all plots, the TLS is taken to be resonant with the bosonic mode $\Omega_d=\Omega_f$.  For all cases, the continuous red line corresponds to $T= 0\ \text{nK}$ and $\gamma = 0\ \text{s}^{-1}$; for the inverted green triangles $T= 5\ \text{nK}$ and $\gamma = 0.5\ \text{s}^{-1}$; finally, for the blue squares $T= 10\ \text{nK}$ and $\gamma = 1\ \text{s}^{-1}$.}
     \label{Fig3}
\end{figure}

To illustrate this transition, we have simulated the system dynamics taking as initial state the qubit excited state and  a thermal state for the phonon field $\rho(0)=\ket{\uparrow}\bra{\uparrow}\otimes\rho_{\rm thermal}$. 
Fig.~\ref{Fig2} shows the average phonon number $N_{\rm ph}$ and Fig.~\ref{Fig3} the population inversion $S_z$ as a function of time, for different values of the coupling strength. As the coupling $g$ is increased, the Jaynes-Cummings evolution is replaced by a less intuitive dynamics. A good parameter to estimate the validity of the RWA is the maximum number of phonons achieved during the system dynamics. When the total phonon number becomes larger than the number of excitations contained in the initial state, the RWA could not provide a good description of the system dynamics. The feasible physical parameters here considered allow to observe the  transition from a dynamics dominated by Rabi oscillations Fig.\ref{Fig2}\textcolor{blue}{(a)} (Fig.\ref{Fig3}\textcolor{blue}{(a)}), to a dynamics that does not preserve the excitation number Fig.\ref{Fig2}\textcolor{blue}{(b)} (Fig.\ref{Fig3}\textcolor{blue}{(b)})
 and, finally, to the fast generation of phonon excitations Fig.\ref{Fig2}\textcolor{blue}{(c)} (Fig.\ref{Fig3}\textcolor{blue}{(c)}).

\subsection{Collapses and revivals in the DSC regime}
When the interaction strength is increased further, and it becomes even larger than the field and TLS frequencies, the system enters the DSC regime \cite{Deep2010}. The most peculiar feature of this regime consists in collapses and revivals of the initial states $\ket{\downarrow,0}(\ket{\uparrow,0})$, which corresponds to initializing the field in the vacuum state and the qubit in the ground (excited) state of its free Hamiltonian term $\hbar\Delta\sigma_z $. Notice that this phenomenon does not correspond to the collapses and revivals observed for coherent states  in the SC regime~\cite{QuantumBook}. Coherent states of the bosonic mode can be feasibly generated with the proposed system \cite{jin1996collective}, and their collapse and revivals can be reproduced and observed. However, here we focus on the DSC regime, which is more challenging to achieve with atomic QED systems.

Collapses and revivals of the vacuum state take place only in the DSC regime, where the interaction term is dominant with respect to the free energy terms~\cite{Deep2010}. Notice that in the USC and DSC regime the system ground state is not the vacuum anymore, but it is given by an entangled state that contains virtual phonons and TLS excitations. This means that the state $\ket{\downarrow,0}$ is actually an excited one, and it is non-trivial to generate in natural implementation of the QRM. In order to initialize the system in such a state, we can take profit of the tunability of the coupling strength provided by the present scheme as it follows. First, the system is initialized  in the strong coupling regime and cooled close to its ground state. Then, the effective interaction strength is suddenly switched to the desired value in the USC or DSC range. To this end, the magnetic field must be changed near a Feshbach resonance, to jump from $g=0$ ($g_{ab}=g_{aa}$) to the values such that $g$ ($g_{ab}>g_{aa}$), in the sub-milisecond regime. In the case of K-Rb mixture, a magnetic field change of less than of 1 G is enough \cite{KRb} to span the interesting range of interaction strength. It has been experimentally proven that such a quench can be achieved in less than 1$\,\operatorname{ms}$ in a controlled fashion \cite{shortquench}.

\begin{figure}[] 
\centering
\includegraphics[width=\linewidth]{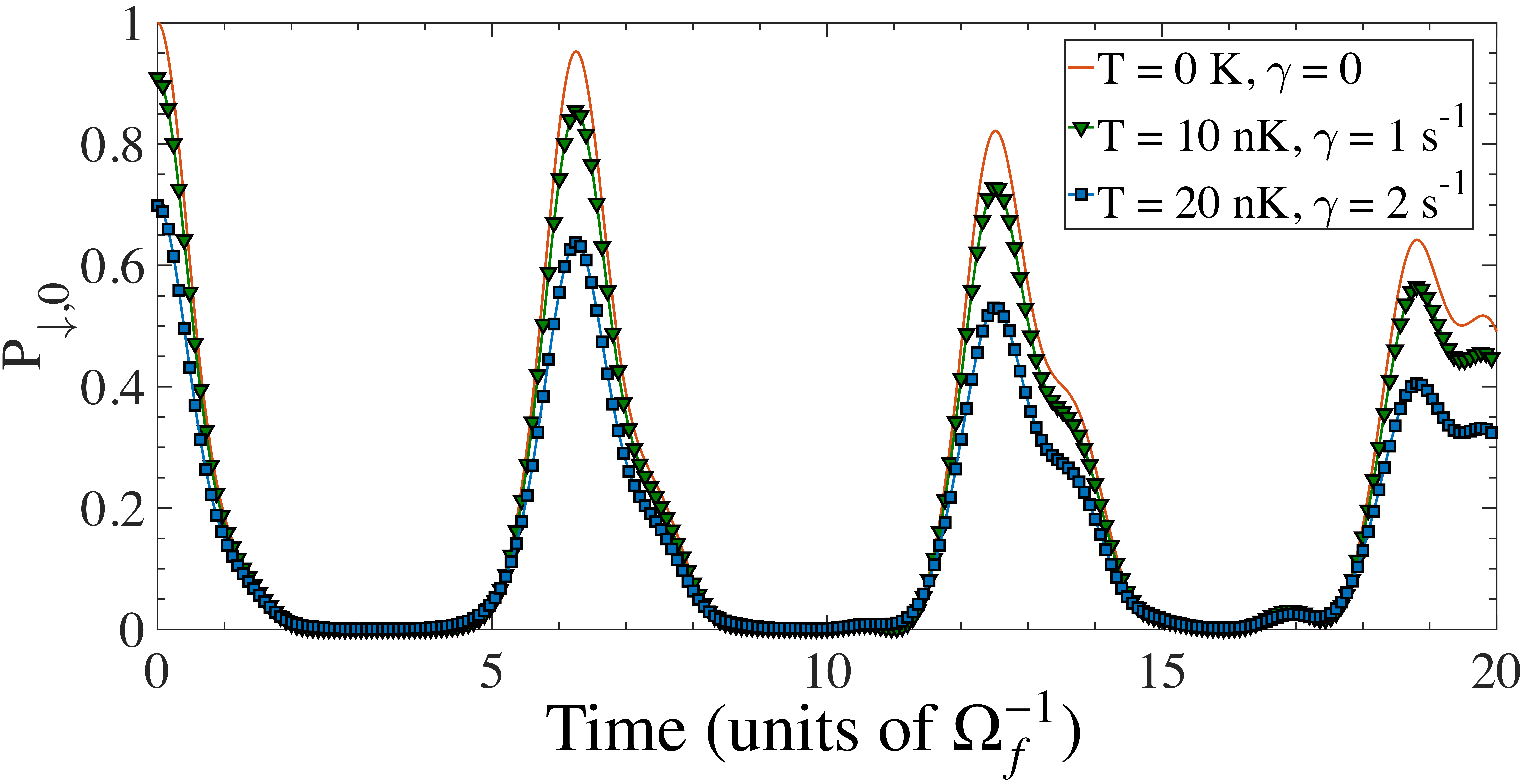}
    \caption{(Color online) Collapse and revival of the probability of finding the TLS in its ground state and the bosonic mode in the vacuum, during system time evolution. The initial state is given by the qubit ground state and a thermal state with temperature $T$ for the bosonic mode. The qubit frequency is given by $\Omega_d=0.1\ \Omega_f$, while for the coupling strength $g=0.8\Omega_f$. The continuous red line corresponds to $T= 0\ \text{nK}$ and $\gamma = 0\ \text{s}^{-1}$; for the green dashed line $T= 10\ \text{nK}$ and $\gamma = 1\ \text{s}^{-1}$, and the blue dashed-dotted line for $T= 20\ \text{nK}$ and $\gamma = 2\ \text{s}^{-1}$.}
     \label{Fig4}
\end{figure}

We have implemented numerical simulations to verify the impact on collapses and revivals of finite temperature effects, that is  imperfect state preparation and amplified dissipation. In Fig.~\ref{Fig4} it is shown the probability $P_{\downarrow,0}(t) = |\langle \downarrow,0|\psi(t)\rangle|^2$ as a function of time and for three different temperatures. The initial condition is chosen as $\rho(0)=\ket{\downarrow}\bra{\downarrow}\otimes\rho_{\rm thermal}$, that is the qubit in its ground state and the phononic mode in a low-excited thermal state.  The dynamics exhibits oscillations coming from the phonon-number wave packet that spread along the parity basis $p=+1$ ($\hat{P}\ket{\downarrow,0}=+\ket{\downarrow,0}$)~\cite{Deep2010}. For all cases we have considered $g/\Omega_f=0.8$, and $\Omega_d/\Omega_f=0.1$. As shown by the blue line with squared markers in Fig.~\ref{Fig4}, collapses and revivals are still visible also at $T= 20\ \text{nK}$.

\section{Effective two-body interactions} 
Let us now extend our model to consider $L$ quantum dots coupled with a single phononic mode. In this case the system Hamiltonian is given by a multi-qubit quantum Rabi model, known as Dicke model in the many-body case,
\begin{equation}
H = \hbar\sum_{n=1}^L \frac{\Omega_{dn}}{2}\sigma^z_n + \hbar\Omega_f a^{\dag}a + \hbar \sum_{n=1}^L g_n\sigma^x_n(a + a^{\dag}).
\label{2RabiH}
\end{equation}
In the dispersive regime, where the qubits are off-resonance with respect to the bosonic mode, the system dynamics results in a effective interactions between the qubits. This condition can be used to implement quantum gates~\cite{Dicarlo09} or quantum simulations of spin chains~\cite{Majer07}.
Here we consider the case where the interaction strength is large enough to break the rotating-wave approximation, even in the dispersive limit~\cite{Zueco2009}. Such interaction leads to an effective Ising-type interaction between atomic quantum dots mediated by a single phononic mode, and single-mode squeezing upon the latter. 
This effective interaction can be obtained from the Hamiltonian~\eqref{2RabiH} by applying the Schrieffer-Wolf transformation $e^{S}He^{-S}$ \cite{SW1966}, with the non-hermitian operator $S=\sum_n[(g_n/\Delta_n)(\sigma_n^{+}a-\sigma_n^{-}a^{\dag}) + (g_n/\delta_n)(\sigma_n^{+}a^{\dag}-\sigma_n^{-}a)]$.
As shown in Ref.\cite{Zueco2009}, the above transformation leads to the effective Hamiltonian
\begin{eqnarray}
H_{\textrm{eff}}&=&\hbar\Omega_fa^{\dag}a+\hbar\sum_{n=1}^L \frac{\Omega_{dn}}{2}\sigma^z_{n} 
\\&+&  \frac{\hbar}{2}\sum_{n=1}^L g_n^2\Bigg(\frac{1}{\Delta_n} + \frac{1}{\delta_n}\Bigg)(a+a^{\dag})^2\sigma^z_{n}\nonumber\\
&+& \frac{\hbar}{2}\sum_{n>m}J_{nm}\sigma^x_{n}\sigma^x_{m}\nonumber,
\label{Heffective}
\end{eqnarray}  
where the effective coupling strengths are defined as 
\begin{equation}
J_{nm}=g_ng_m\Bigg(\frac{1}{\Delta_n}+\frac{1}{\Delta_m}-\frac{1}{\delta_n}-\frac{1}{\delta_m}\Bigg),
\end{equation}
and $\Delta_n=\Omega_{dn}-\Omega_f$ and $\delta_n=\Omega_{dn}+\Omega_f$.

\begin{figure}[t] 
\centering
\includegraphics[width=0.9\linewidth]{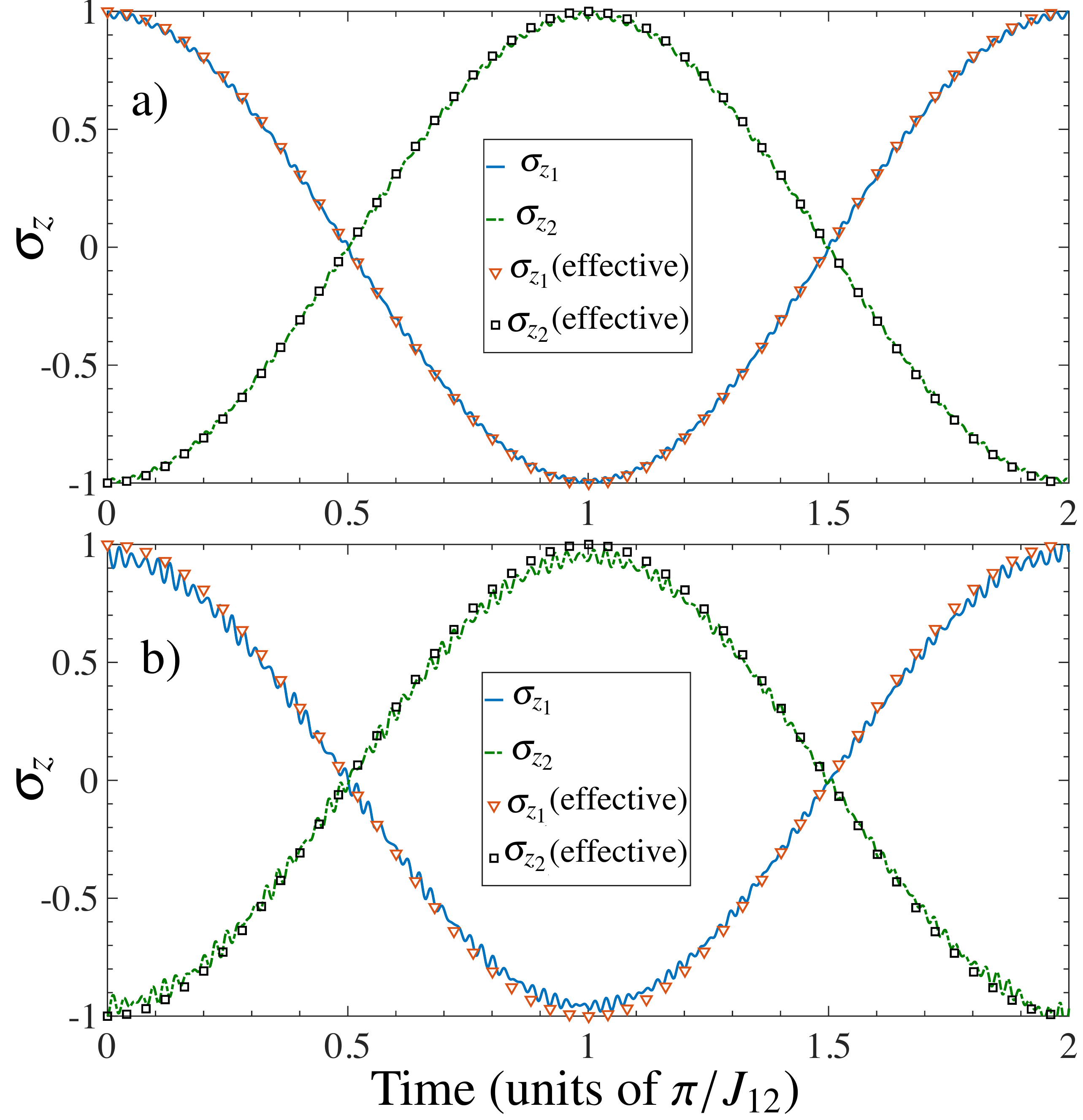}
    \caption{(Color online) Expected value of $\sigma_z$ along time evolution. a) $T = 5~\textrm{nK}$ and b) $T = 100~\textrm{nK}$. In all plots we consider $\Omega_f=2\pi\times10^3\textrm{s}^{-1}$, $\gamma=1\textrm{s}^{-1}$, and identical TLSs with parameters $\Omega_{d1}=\Omega_{d2}=0.1~\Omega_f$, and coupling strength $|g|=0.054~\Omega_f$. For all cases, the continuous blue line and the green dot-dashed lines correspond to the average value of $\sigma_z$ for the first and second TLS respectively, when the full Hamiltonian~\eqref{2RabiH} is considered. Inverted triangles and square markers correspond to the same observables, for the dynamics of the effective model~\eqref{Heffective}. }
     \label{Fig5}
    \end{figure}

We have performed numerical simulations to check the validity of this approximation at finite temperature. 
In Fig.~\ref{Fig5} we show the results obtained comparing the dynamics of the full~\eqref{2RabiH} and effective~\eqref{Heffective} Hamiltonian.
The plots show the  exchange of excitations between two identical qubits as function of time, when the initial condition is given by $\rho(0)=\ket{\uparrow}\bra{\uparrow}\otimes\ket{\downarrow}\bra{\downarrow}\otimes\rho_{\rm thermal}$, for different temperatures of the bosonic mode. We find that the effective Hamiltonian describes accurately the system dynamics for feasible low temperatures [see Fig.~\ref{Fig5}(a)]. When significant thermal effects are included, as in Fig.~\ref{Fig5}(b), 
small oscillation at a fasters timescale appear, but the effective model still provides a qualitative description of the excitation exchange.

\section{Conclusions} 
We  propose a novel scheme to implement a quantum simulation of the quantum Rabi model in a cold-atom setup. It consists of an atomic quantum dot coupled to the quasiparticle excitations of a BEC with suitable hard-wall boundary conditions. By exploiting the dependence of the atomic collision strengths on external magnetic fields near a Feshbach resonance, the parameters of the condensate can be tuned over a wide interval spanning the SC, USC and DSC regime of interaction. An important feature of our setup is the slow propagation speed of the quasiparticles whose characteristic frequencies are in the  of tens or hundreds of Hz. This results in a slow time scale that facilitates the experimental preparation of the initial states and the control of the interaction strength required to analyze the features of the quantum Rabi model. Our scheme can be naturally extended including additional two-level systems. We have analyzed the appearance of effective phonon-mediated pairwise interactions. Our scheme would allow to experimentally investigate  the effect of finite temperature on dispersively-coupled qubits in the transition between the strong and the ultrastrong coupling regimes.

\section*{Acknowledgements} S.F. acknowledges support from the PRESTIGE program, under the Marie Curie Actions-COFUND of the FP7. G.R. acknowledges the support from the Fondo Nacional de Desarrollo Cient\'ifico y Tecnol\'ogico (FONDECYT, Chile) under grant No. 1150653.  E.S. acknowledges financial support from  Spanish MINECO/FEDER FIS2015-69983-P and Basque Government IT986-16. Financial support by Fundaci{\'o}n General CSIC (Programa ComFuturo) is acknowledged by C.S as well as additional support from Spanish MINECO/FEDER FIS2015-70856-P and CAM PRICYT Project QUITEMAD+ S2013/ICE-2801.

\bibliographystyle{apsrev}
\bibliography{biblio.bib}


\end{document}